\documentclass[%
 reprint,
amsmath, amssymb,
aps,
]{revtex4-2}

\usepackage{verbatim}  

\usepackage{hyperref} 
\hypersetup{
    colorlinks=true,
    linkcolor=blue,
    filecolor=blue,      
    urlcolor=blue,
    pdftitle={OverlappingContagion},
    pdfpagemode=FullScreen,
    }
\usepackage{amsmath}  
\usepackage{amssymb}  
\usepackage{amsthm}  
\usepackage{mathtools}  
\usepackage{bbm}  

\usepackage{cleveref}[capitalise]

\newcommand{\change}[1]{#1}

\begin{document}
\title{Reconstruction of multiplex networks via graph embeddings}

\author{Daniel Kaiser}
\author{Siddharth Patwardhan}
\author{Minsuk Kim}
\author{Filippo Radicchi}
    \email{filiradi@indiana.edu}
\affiliation{
    Center for Complex Networks and Systems Research, Luddy School of Informatics, Computing, and Engineering\\
    Indiana University, Bloomington, Indiana 47408, USA
}

\date{\today}

\begin{abstract}
    {
        Multiplex networks are collections of networks with identical nodes but distinct layers of edges. 
        They are genuine representations for a large 
        variety of real systems whose elements interact in multiple fashions or flavors.
        However, multiplex networks are not always simple to observe in the real world;
        often, only partial information on the layer structure of the networks is available, whereas the remaining information is in the form of aggregated, single-layer networks. 
        Recent works have proposed solutions to the problem of reconstructing the hidden multiplexity of single-layer networks using tools proper of network science. 
        Here, we develop a machine learning framework that takes advantage of graph embeddings, i.e., representations of networks in geometric space. 
        We validate the framework in systematic experiments aimed at the reconstruction of synthetic and real-world multiplex networks, providing evidence that our proposed framework not only accomplishes its intended task, but often outperforms existing reconstruction techniques. 
    }
\end{abstract}

\maketitle

\section{Introduction}
It has been shown several times over that applying classical network models to systems with several dimensions of interactions is an unfaithful representation of reality, often dangerously so~\cite{boccalettiStructureDynamicsMultilayer2014b,dongRobustnessNetworkNetworks2013,gaoRobustnessNetworkNetworks2011a,zaninCanWeNeglect2015,bianconi2018multilayer,osat2017optimal}.
Examples range from cascading power outages~\cite{buldyrevCatastrophicCascadeFailures2010, liuBreakdownInterdependentDirected2016a, leeThresholdCascadesResponse2014}
and models of the brain~\cite{battistonMultilayerMotifAnalysis2017a,dedomenicoMultilayerModelingAnalysis2017,mandkeComparingMultilayerBrain2018,vaianaMultilayerBrainNetworks2020}, 
to efficient transportation with several modalities~\cite{ferrariVulnerabilityRobustnessInterdependent2023,wuTrafficDynamicsMultilayer2020,dedomenicoNavigabilityInterconnectedNetworks2014a,gallotti2014anatomy,strano2015multiplex} 
and efficient epidemic mitigation~\cite{chinazziEffectTravelRestrictions2020,sanhedraiSustainingNetworkControlling2023,alvarez-zuzekDynamicVaccinationPartially2019}.
Problematically, it is sometimes the case that only a monoplex of interactions can be observed; perhaps with some partial, incomplete observations of the truly multidimensional ways in which some entities interact. Hence, there is want of a method to infer likely multiplexes from these incomplete observations. Researchers have only recently begun to approach this problem which we shall broadly call the multiplex reconstruction problem (MRP).

Several approaches to the MRP have been considered so far.
Tarr\'es-Deulofeu {\it et al.}~\cite{tarres-deulofeuTensorialBipartiteBlock2019} and De Bacco {\it et al.}~\cite{debaccoCommunityDetectionLink2017a} utilize statistical inference techniques grounded in the tensorial formulation of multilayer networks by De Domenico {\it et al.}~\cite{dedomenicoMathematicalFormulationMultiLayer2013}. Their work allows for attacking the more classical link prediction task in multiplexes with a well-defined block-model-esque generative prior, potentially even including layer-wise correlation. While very general and quite powerful, both methods are designed with link prediction in the usual sense in mind, hence, utilize training sets that are significantly larger than the test set of edges to actually predict. They consequently suffer from inefficiencies and biases in the face of sparse \textit{a priori} information ~\cite{tarres-deulofeuTensorialBipartiteBlock2019}.
Zhang {\it et al.}~\cite{zhangDetanglingMultilayerStructure2021} instead 
consider two-layer multiplex networks and do not necessitate a notion of partial observations. Their work employs a simulated annealing algorithm to reconstruct layers maximizing an objective function inspired by clustering measures, hence, they are able to reconstruct multiplex networks with sufficiently layer-endogenous motifs. Similarly to other approaches, Zhang {\it et al.}'s work also suffers some in the face of edge sparsity~\cite{zhangDetanglingMultilayerStructure2021}.
Kaiser {\it et al.}~\cite{kaiser2023multiplex} and Wu {\it et al.}~\cite{wu2022discrimination} both leverage partial observations of the underlying true multiplex structure that may be available \textit{a priori} along with assumptions on a layer-wise generative model of the system grounded in the layer-wise degree sequences. Wu {\it et al.} utilize an expectation-maximization framework to approximate the layer-wise degree sequence ~\cite{wu2022discrimination}, whereas Kaiser {\it et al.} construct an approximating multiplex network from which they extract degree sequences ~\cite{kaiser2023multiplex}. Both techniques, then, are able to use their generative model to sample likely reconstructions of the true multiplex structure.
Also these approaches have limitations. Wu {\it et al.}'s technique makes no use of mesoscale structure, instead relying entirely on global degree sequence structure~\cite{wu2022discrimination}. Kaiser {\it et al.} improve on this insofar as considering assortative block structure at all in their work; the techniques they leverage are able to be confounded by improperly assuming assortative block structure where they may be none, however ~\cite{kaiser2023multiplex}. If the system attempting to be reconstructed is sufficiently unknown, then it might be difficult to assess if there is assortative community structure in the true multiplex 
that one is attempting to reconstruct. 

In this work we propose a methodology of reconstructing multiplex networks that accounts for both global degree sequence and mesoscale block structure information; furthermore, our proposed methodology is able to regulate between these two sources of information without 
the need of introducing too much prior knowledge about the system.
Our approach is inspired by the work of Kaiser {\it et al.}~\cite{kaiser2023multiplex}, albeit with an added focus on the needed robustness and flexibility. We design a similar reconstruction algorithm centered around calculating the likelihood of an edge in the observed aggregate network of a multiplex to have originated from a particular layer of the underlying ``true" multiplex. As in both Kaiser {\it et al.} and Wu {\it et al.}, in the absence of further information, the principle of parsimony guides us to assume a configuration model-esque likelihood on the basis of multi-degree sequence~\cite{kaiser2023multiplex,wu2022discrimination}. We extend this likelihood to include geometric representation of mesoscale structure, however, by further utilizing graph embeddings to encode this structure. The rationale of using an embedding-aided approach to the MRP naturally stems from the well-established usefulness of these graph representations in the task of predicting missing links in single-layer networks~\cite{cai2018comprehensive, goyal2018graph, hamilton2017representation}.  In our method, we calculate two sources of information that contribute to each edge's likelihood to have originated from a particular layer of the underlying multiplex, namely, multi-degree sequence under a generative model assumption and distance in a latent embedding space. In order to flexibly utilize these data, we train a logistic regression model on suitably normalized features and retrieve a distribution of multiplex reconstructions. 

We examine the resultant reconstructed multiplex networks in this work and explore the potency of the proposed methodology. That is, we apply the developed method to a corpus of multiplex systems consisting of both synthetic models and real-world systems. In our investigations on synthetic models, we push the method to its limits and characterize structures that are more or less capable of being reconstructed. When applying the method to real-world systems we find comparable reconstruction performance of current methods but with significantly more stable reconstructions. Furthermore, we are able to reconstruct systems where previous methods struggle and are confounded by mesoscale structure.

\section{Methods}\label{sec:methods}
\subsection{The multiplex reconstruction problem}\label{sec:methods:mrp}
We consider multiplex networks with two layers, namely $\alpha$ and $\beta$. The graph representing layer $\alpha$ is denoted by $\mathcal{G}^{(\alpha)} = \{ \mathcal{N}^{(\alpha)},  \mathcal{E}^{(\alpha)}\}$, where $\mathcal{N}^{(\alpha)}$ is the set of nodes of the layer, and $\mathcal{E}^{(\alpha)}$ is the set of its edges. 
$\mathcal{N}^{(\beta)}$  and $\mathcal{E}^{(\beta)}$ are respectively the sets of nodes and edges of layer $\beta$. 
Being a multiplex, we assume that $\mathcal{N} = \mathcal{N}^{(\alpha)} = \mathcal{N}^{(\beta)}$, and we indicate the size of the network with $N = \left|\mathcal{N}\right|$. 
We exclude the possibility that the edge $(i,j)$ belongs to both layer $\alpha$ and $\beta$ simultaneously, i.e., $\mathcal{E}^{(\alpha)} \cap \mathcal{E}^{(\beta)} = \emptyset$. We define the multiplex reconstruction problem (MRP) as the binary classification of the individual edges in the multiplex layers, i.e.,  predicting whether the generic edge $(i,j)$ belongs to either  $\mathcal{E}^{(\alpha)}$ or $\mathcal{E}^{(\beta)}$ \cite{kaiser2023multiplex}. 

Similar to Refs.~\cite{kaiser2023multiplex} and~\cite{wu2022discrimination}, we perform the classification using partial knowledge of the multiplex network. We assume to know the elements of the union $\mathcal{E}^{(\alpha)} \cup \mathcal{E}^{(\beta)}$; however, we also assume to know the layer which individual edges belong to only for a fraction of them, namely the training sets $\mathcal{E}_{\textrm{train}}^{(\alpha)} \subseteq \mathcal{E}^{(\alpha)}$ and $\mathcal{E}_{\textrm{train}}^{(\beta)} \subseteq \mathcal{E}^{(\beta)}$, respectively. We train a classification model on these sets and we use it to classify all edges $(i,j) \in  \mathcal{E}_{\textrm{test}}^{(\alpha)} \cup \mathcal{E}_{\textrm{test}}^{(\beta)}$, with $\mathcal{E}_{\textrm{test}}^{(\alpha)} =\mathcal{E}^{(\alpha)} \setminus \mathcal{E}_{\textrm{train}}^{(\alpha)}$ and $\mathcal{E}_{\textrm{test}}^{(\beta)} =\mathcal{E}^{(\beta)} \setminus \mathcal{E}_{\textrm{train}}^{(\beta)}$. We measure the classifier's performance using the area under the receiver operating characteristic curve (ROC-AUC), i.e., a standard metric in binary classification tasks.

Results of the classification model proposed in this paper are contrasted to those obtained using the degree- and community-based classifier (DC) by Kaiser {\it et al.}~\cite{kaiser2023multiplex}. The DC classifier takes advantage of the community structure of the multiplex network to predict the flavor of its edges. Virtually, any community detection algorithm can be used in the DC classifier. Results reported in this paper are based on communities detected using the Louvain algorithm~\cite{blondel2008fast}.

\subsection{Classification model}\label{sec:methods:model}

\begin{figure*}[!tb]
    \centering
    \includegraphics[width=0.9\textwidth]{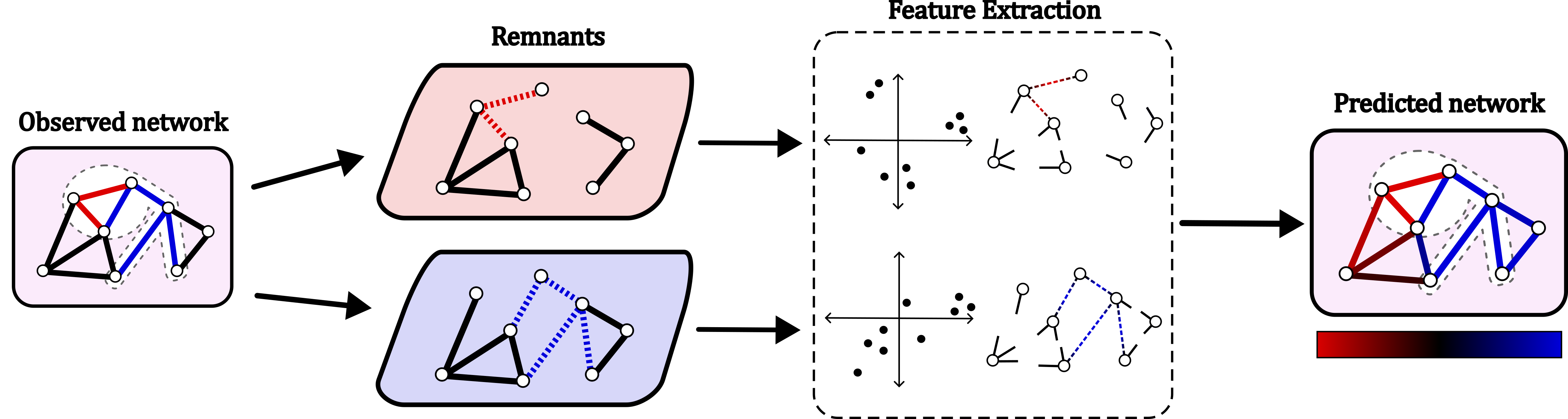}
    \caption{
        {\bf Schematic of the proposed workflow for multiplex reconstruction.} Beginning from an observed single-layer network with partial observation of the true dimensionality of some edges, we first construct the remnant networks. We then extract features from the constructed remnants to inform a classification model. Here, the features we extract are functions of the distances of embedded nodes incident to each edge in each remnant layer 
        [i.e., \Cref{eq:emb}]
        as well as the product of degrees in each remnant layer 
        [i.e., \Cref{eq:degree}]. 
        We also extract an edge-balance  feature 
        [i.e., \Cref{eq:balance}] 
        that, however, serves only to properly establish the offset of the classifier. From these features, we train a logistic regression model to classify each edge within one of the two layers 
        [i.e., \Cref{eq:logistic}].
        The result of the reconstruction workflow is a weighted network where the weight of each edge indicates its likelihood to have originated from a particular layer. 
    }
    \label{fig:schematic}
\end{figure*}

We illustrate the reconstruction workflow proposed in this article in \Cref{fig:schematic}. All code is available at~\cite{repo}. 
The input of the workflow is a partially observed multiplex network. Outputs are probabilities for individual, non-observed edges to belong to either one or the other layer of the multiplex network.

In the framework, we employ a logistic regression model
that relies on features defined for
the remnant networks of the available partial information, i.e., the networks obtained from the edges belonging to the test set, but not in the training set of the other layer. Formally, the remnant networks for layers $\alpha$ and $\beta$ are respectively given by $\mathcal{G}_{R}^{(\alpha)} = \{ \mathcal{N}, \mathcal{E}_{\textrm{test}} \setminus \mathcal{E}_{\textrm{train}}^{(\beta)} \}$ and $\mathcal{G}_{R}^{\beta} = \{ \mathcal{N}, \mathcal{E}_{\textrm{test}} \setminus \mathcal{E}_{\textrm{train}}^{(\alpha)} \}$, where $ \mathcal{E}_{\textrm{test}}= \mathcal{E}_{\textrm{test}}^{(\alpha)} \cup \mathcal{E}_{\textrm{test}}^{(\beta)}$. 

First, we compute the edge-balance feature as
\begin{equation}
\zeta = 2 \, \frac{| \mathcal{E}_{\textrm{train}}^{(\alpha)}|}{| \mathcal{E}_{\textrm{train}}^{(\alpha)}| + | \mathcal{E}_{\textrm{train}}^{(\beta)}|} - 1 \; .
\label{eq:balance}
\end{equation}
The above feature is based on the assumption that observed edges are randomly selected from the set
$\mathcal{E}^{(\alpha)} \cup \mathcal{E}^{(\beta)}$. We have $-1 \leq \zeta \leq 1$. Positive values of $\zeta$ indicate propensity of an edge to be classified in layer $\alpha$, whereas negative values of $\zeta$ indicate that the edge is more likely to belong to layer $\beta$.

Second,
we associate to each edge
$(i,j)$ the score 
\begin{equation}
\kappa_{i,j} = 2 \, \frac{k_i^{(\alpha)} k_j^{(\alpha)}}{k_i^{(\alpha)} k_j^{(\alpha)} + k_i^{(\beta)} k_j^{(\beta)}} - 1
\label{eq:degree}
\end{equation}
where $k_i^{(\alpha)}$ denotes the degree of node $i$ in the remnant network for layer $\alpha$, and similarly for $k_i^{(\beta)}$. 
We have $-1 \leq \kappa_{i,j} \leq 1$. Positive values of $\kappa_{i,j}$ indicate propensity of the edge $(i,j)$ to be classified in layer $\alpha$, whereas negative values of $\kappa_{i,j}$ indicate that the  edge $(i,j)$ is more likely to belong to layer $\beta$. Except for a linear transformation, this is the same score already considered in the multiplex classifiers by Wu {\it et al.}~\cite{wu2022discrimination} and Kaiser {\it et al.}~\cite{kaiser2023multiplex}.

Finally,
given the remnants $\mathcal{G}_{R}^{(\alpha)}$ and $\mathcal{G}_{R}^{(\beta)}$, each node $i$ in the graph is mapped to two points $\mathbf{x}^{(\alpha)}_i$ and $\mathbf{x}^{(\beta)}_i$, respectively, into $d$-dimensional Euclidean spaces. Both layer-wise mappings rely on the same embedding technique, but they are performed 
separately,
meaning that the mapping $\mathbf{x}^{(\alpha)}_i$ uses only information from the graph $\mathcal{G}_{R}^{(\alpha)}$ and the mapping $\mathbf{x}^{(\beta)}_i$ uses only information from the graph $\mathcal{G}_{R}^{(\beta)}$. Once the embeddings $\mathbf{x}^{(\alpha)}_i$ for all nodes $i$ are performed, we manipulate them as follows. We 
transform 
the embedding vectors in each connected component $\mathcal{C}^{(\alpha)} \subseteq \mathcal{G}_R^{(\alpha)}$ 
such that
\begin{equation}
\sum_{i \in \mathcal{C}^{(\alpha)}} \, \mathbf{x}_{i}^{(\alpha)}  = \mathbf{0} \; ,
    \label{eq:scale}
\end{equation}
where $\mathbf{0}$ is the vector with components all equal to zero, 
and then renormalize them
such that
\begin{equation}
\sum_{i \in \mathcal{C}^{(\alpha)}} \, \lVert\mathbf{x}_{i}^{(\alpha)}\rVert  = 1 \; ,
    \label{eq:norm}
\end{equation}
where $\lVert \mathbf{x} \rVert$ is the $\text{L}_2$-norm of the vector $\mathbf{x}$.
A similar procedure is applied also to the embeddings 
$\mathbf{x}^{(\beta)}_i$  obtained from the graph $\mathcal{G}_{R}^{(\beta)}$. Finally, for any edge $(i,j)$, we define the score
\begin{equation}
\delta_{i,j} = 2 \, \frac{\lVert\mathbf{x}_i^{(\alpha)} - \mathbf{x}_j^{(\alpha)}\rVert^{-1}}{\lVert\mathbf{x}_i^{(\alpha)} - \mathbf{x}_j^{(\alpha)}\rVert^{-1} + \lVert\mathbf{x}_i^{(\beta)} - \mathbf{x}_j^{(\beta)}\rVert^{-1}} - 1\; ,
\label{eq:emb}
\end{equation}
where $\lVert\mathbf{x} - \mathbf{x}'\rVert$ is the Euclidean distance between the points $\mathbf{x}$ and $\mathbf{x}'$ as long as $\mathbf{x} \neq \mathbf{x}'$; we set instead $\lVert\mathbf{x} - \mathbf{x}'\rVert = 2^{-32}$ if $\mathbf{x} = \mathbf{x}'$ to avoid numerical overflow.
Also here, we have $-1 \leq \delta_{i,j} \leq 1$. Positive values of $\delta_{i,j}$ indicate propensity of the edge $(i,j)$ to be classified in layer $\alpha$, whereas negative values of $\delta_{i,j}$ indicate propensity of the edge $(i,j)$ to be classified in layer $\beta$.

We note that each connected component is mapped independently from the others, hence
the per-component transformations described in 
\Cref{eq:norm,eq:emb}
serve to suppress eventual dependencies of the embeddings on the components' sizes.

Finally, we train a logistic regression model using the edges in the training set
$\mathcal{E}^{(\alpha)}_{\text{train}} \cup \mathcal{E}^{(\beta)}_{\text{train}}$. We rely on the Python library scikit-learn \cite{scikit-learn}. 
The model allows us to estimate for each edge $(i,j) \in \mathcal{E}_{\text{test}}$ the likelihood of originating from layer $\alpha$ as

\begin{equation}
        P \big( (i,j) \in \mathcal{E}^{(\alpha)} \big) = 
        \frac{1}{1 + \exp{\left[-(a \zeta + b \delta_{i,j} + c \kappa_{i,j}) \right]}}
        \; .
        \label{eq:logistic}
\end{equation} 
The coefficients $a$, $b$ and $c$ weigh the importance of the various features in the classification task. Specifically, $a$ proxies the importance of the balance between the sizes of the training sets $\mathcal{E}^{(\alpha)}_{\text{train}}$ and $\mathcal{E}^{(\beta)}_{\text{train}}$;
$b$ and $c$ respectively quantify the relevance of the features $\kappa$ and $\delta$ in the reconstruction problem. 

The complexity of the necessary calculations in \Cref{eq:balance,eq:degree,eq:scale,eq:norm,eq:emb,eq:logistic} are dominated by the chosen graph embedding method. We report asymptotic time complexity of our workflow in \Cref{sec:appendix:time}.

\subsection{Graph embedding techniques}\label{sec:methods:embeddings}
We consider four popular methods to embed the remnants $\mathcal{G}_R^{(\alpha)}$ and $\mathcal{G}_R^{(\beta)}$ of a multiplex network in Euclidean space: Node2Vec (N2V)~\cite{grover2016node2vec}, a modified version of Laplacian eigenmaps (LE)~\cite{belkin2001laplacian}, 
high-order proximity preserved embedding (HOPE)~\cite{ou2016asymmetric}, and Isomap~\cite{tenenbaum2000global}. Below, we provide some brief descriptions of these methods, and report on the choice of the parameters that we used to perform the graph embeddings.

In all methods, the dimension $d$ of the embedding space is a free parameter. We fix $d = \min \{N, 128 \}$ throughout this article.

\subsubsection{Node2Vec}\label{sec:methods:embeddings:n2v}
N2V \cite{grover2016node2vec} involves creating multiple node sequences through random walks of fixed length and then optimizing the representations of the nodes in the Euclidean space to maximize the likelihood of node co-occurrence in the sequences. N2V has several tunable parameters including walk length, window size, bias parameters $p$ and $q$ for the random walk, and the embedding dimension $d$. Throughout the paper, we set the window size to be $10$, the walk length to be $80$, and $p = q = 1$. We take advantage of the systematic analysis of embedding methods conducted in Ref.~\cite{zhang2021systematic} to appropriately choose these values of the parameters.

\subsubsection{Laplacian eigenmaps}\label{sec:methods:embeddings:le}
LE \cite{belkin2001laplacian} 
is an embedding method based on the spectrum of the normalized Laplacian operator of the graph to be embedded. Indicate with $0 < \lambda_1 \leq \ldots \leq \lambda_r \leq \ldots \leq \lambda_d$ the $d$ smallest, non-trivial eigenvalues of the normalized Laplacian, and with $\mathbf{v}_1, \ldots, \mathbf{v}_r, \ldots, \mathbf{v}_d$ the corresponding eigenvectors. The standard implementation of LE maps node $i$ to point $\mathbf{x}_i = (v_{1,i}, \ldots, v_{r,i}, \ldots, v_{d,i})$, where $v_{r,i}$ is the $i$-th component of the  eigenvector associated with $\lambda_r$.  In this paper, we modify the standard LE embedding by suppressing the contribution of each eigenvector by the value of its corresponding eigenvalue. We basically embed node $i$ as $\mathbf{x}_i = (v_{1,i}/\lambda_1, \ldots, v_{r,i}/\lambda_r, \ldots, v_{d,i}/\lambda_d)$.
\change{The modification allows to properly weigh the importance of the various modes of the normalized Laplacian operator to represent the eventual modular structure of the network layer. We empirically verified that such a modification greatly improves the performance in the downstream reconstruction problem.}

\subsubsection{HOPE}\label{sec:methods:embeddings:hope}
HOPE~\cite{ou2016asymmetric} tries to preserve a given similarity matrix $S$ in the Euclidean space of the desired dimension $d$ by minimizing

\begin{equation}
    E_{H} = \sum_{i, j}\lVert S_{ij} - \mathbf{x}^{T}_{i}\mathbf{x}_{j} \rVert \; .
\end{equation}
 
Although HOPE has the flexibility to utilize various node similarity matrices, we adopt the common practice of using the Katz index in this work.

\subsubsection{Isomap}\label{sec:methods:embeddings:isomap}
The objective function in Isomap aims to preserve the pairwise shortest-path distances between nodes \cite{tenenbaum2000global}. The shortest path-distance matrix $\mathbf{D}$ is computed for a network in the first step. Then multidimensional scaling is applied to $\mathbf{D}$ to obtain a vector representation of the nodes in the Euclidean space of desired dimension $d$ that minimizes the following objective function
\begin{equation}
    E_I = \sum_{i,j} \big(D_{ij} - \lVert \mathbf{x}_i - \mathbf{x}_j \rVert\big)^2 \; .
\end{equation}

\section{Data}\label{sec:data}
\subsubsection{Synthetic multiplex networks}\label{sec:data:synth}

We generate network layers using the Lancichinetti-Fortunato-Radicchi (LFR) model~\cite{lancichinetti2008benchmark}.
We begin by generating an instance of the LFR model with a given set of parameters. We fix the value of the community size power-law exponent $\tau = 1.0$ and the maximum degree $k_{\max} = \sqrt{N}$ for both layers. We vary the degree exponent $\gamma$ and the mixing parameter $\mu$, although we consider only experiments where these parameters are identical for both layers. Instead, we vary the average degree $\langle k^{(\alpha)} \rangle$ and $\langle k^{(\beta)} \rangle$ at the level of the individual layers. We do not impose any constraints on the size and number of communities. 
The two network layers are generated independently, thus edge overlap and correlation among the layer-wise community structures are negligible. If any edge is shared by the two layers, we remove it from the system.

\subsubsection{Real multiplex networks}\label{sec:data:real}
We analyze several real-world multiplex networks. Even if some datasets consist of more than two layers, our tests are performed considering two layers at a time and all possible pairings of the layers. For each combination of the layers, edges shared by both layers are deleted and no information on their existence is considered in the MRP. Then, the set of nodes in the corresponding multiplex is given by the union of the sets of nodes of the two individual layers. Details on real datasets analyzed beyond those present in the figures are reported in Table~\ref{tab:summary_real_multiplexes_pairs}. 

\begin{table*}[!hbt]
    \centering
    \begin{tabular}{|l||r|r|r|r||r|r|r|r|r|}
        \hline
        Dataset & $N$ & $|\mathcal{E}^{(\alpha)}|$ &  $|\mathcal{E}^{(\beta)}|$ & $|\mathcal{E}^{(\alpha, \beta)}|$  & DC & HOPE & Isomap & LE & N2V \\
        \hline
        \hline
        \textit{R. norvegicus}~\cite{de2015structural} & 2520 & 2610 & 832 
        & 178
        & 0.69 & 0.71 & 0.72 & 0.70 & \textbf{0.73}\\

        \hline
        \textit{S. pombe}~\cite{de2015structural} & 2577 & 1038 & 6622 
        & 445
        & \textbf{0.67} & 0.63 & 0.62 & 0.61 & 0.65\\

        \hline
        \textit{H. sapiens}~\cite{de2015structural} & 17415 & 37361 & 72004 
        & 12284 
        & 0.91 & 0.91 & 0.91 & 0.92 & \textbf{0.93}\\
        
        \hline
        European airlines~\cite{de2015structural} & 128 & 244 & 601
        & 0
        & 0.92 & 0.94 & 0.94 & 0.94 & \textbf{0.95}\\
        
        \hline
    \end{tabular}
    \caption{
        {\bf Summary statistics of real-world multiplex networks.} From left to right, we report the name of the dataset and reference where the dataset was introduced, the number of nodes within the two layers,  the number of edges for each of the two layers, the number of edges shared by the two layers, and the average ROC-AUC of reconstruction at 30\% training set size of Kaiser {\it et al.}'s previous reconstruction method~\cite{kaiser2023multiplex} and our proposed method. The reconstruction method with best average performance for each multiplex is bolded among five repetitions. For the genetic interaction systems (italicized), we consider the layers for direct and suppressive gene interactions. For the European Airlines system, we consider the airlines for the companies Lufthansa and Ryanair.
    }
    \label{tab:summary_real_multiplexes_pairs}
\end{table*}

\section{Results}\label{sec:results}
\subsection{Synthetic networks}\label{sec:results:synth}
\subsubsection{Performance on synthetic networks}
We begin by seeking to understand the fundamental capabilities and limitations of the proposed framework. To this end, we sample synthetic multiplexes as described in \Cref{sec:data:synth} and conduct a systematic investigation. Our intention is to elucidate how performant the proposed reconstruction workflow can be in scenarios with known assortative community structure and degree distribution, hence, to explore how applicable multiplex reconstruction is in broad strokes.

Using the LFR benchmarks, we consider four possible scenarios that can
capture the structure of real-world systems:
\begin{itemize}
    \item Strong communities and heterogeneous degrees, i.e.,  $\mu=0.1$ and $\gamma=2.1$;
    \item Strong communities and homogeneous degrees, i.e.,  $\mu=0.1$ and $\gamma=4.0$;
    \item Mediocre communities and moderately heterogeneous degrees, i.e.,  $\mu=0.3$ and $\gamma=2.7$;
    \item Weak communities and heterogeneous degrees, i.e., $\mu=0.5$ and $\gamma=2.1$.
\end{itemize}
In addition to these parametrizations of mesoscale-global structure, we 
fix the network size to $N=10,000$ and the average degrees of the layers $\langle k^{(\alpha)} \rangle = \langle k^{(\beta)} \rangle = 6$.
We then apply the proposed reconstruction methods to samples of these models and quantify the reconstruction performance as a function of the size of the training set. See \Cref{sec:appendix:size} for more on the effect of network size on reconstruction. 
 Please note that the experimental condition $\langle k^{(\alpha)} \rangle = \langle k^{(\beta)} \rangle$ immediately implies that the balance feature $\zeta \simeq 0$, thus we also have that $a \simeq 0$ in the logistic regression model of 
 \Cref{eq:logistic}.

\begin{figure}[htb]
    \centering
    \includegraphics[width=0.47\textwidth]{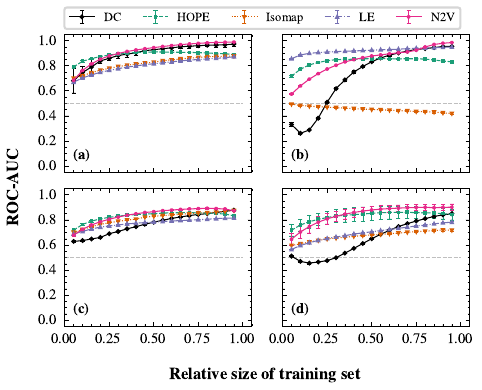}
    \caption{
        {\bf Reconstruction of synthetic multiplex networks.}
        Each panel depicts the performance as a function of training information available on LFR duplexes with fixed parameters $N=10,000, \langle k^{(\alpha)} \rangle = \langle k^{(\beta)} \rangle = 6, k_{max} = 100, \tau = 1.0$. We vary $\mu, \gamma$ across panels as: (a) $\mu=0.1, \gamma=2.1$; (b) $\mu=0.1, \gamma=4.0$; (c) $\mu=0.5, \gamma=2.1$; (d) $\mu=0.3, \gamma=2.7$. Different combinations of colors and symbols correspond to different reconstruction methods. The horizontal dashed line indicates the performance value of a random classifier.
        All points depict the mean of ten realizations with error bars displaying the standard error. 
    }
    \label{fig:perfs-LFR}
\end{figure}

Several interesting behaviors are immediately observable in the reconstruction performance as illustrated in \Cref{fig:perfs-LFR}, where each figure panel corresponding to a different scenario as described above. Examining N2V's performance first, it is clear that, regardless of the values of $\mu \text{ and } \gamma$ in these panels, N2V-based reconstruction consistently performs well; with especially notable performance in \Cref{fig:perfs-LFR} panels (a) and (c) where there is strong degree heterogeneity, outperforming every other method for a majority of training sets. Indeed, \Cref{fig:perfs-LFR} is highly suggestive of the stability in the performance of N2V. While generally second to N2V, except at small sizes of the training set, HOPE also displays potent consistency across the varied scenarios examined here. 

LE is able to surpass the other methods only in the case of \Cref{fig:perfs-LFR} panel (b) where there is strong degree homogeneity. Otherwise, both LE and Isomap are generally poor choices for reconstruction. Especially concerning is the performance of Isomap where there is a strong community structure but exceptionally homogeneous degrees as seen in \Cref{fig:perfs-LFR} panel (b). Here we see the only case of a reconstruction method consistently performing \textit{worse} than a random classifier. A random binary classifier on balanced classes should yield an ROC-AUC $\simeq 0.5$, as shown by the dashed gray line. Isomap's lesser performance than this guideline may not be entirely unexpected, however, as Isomap generally does not perform well for graph clustering.

Perhaps the most notable comparison in \Cref{fig:perfs-LFR} is against the DC reconstruction method by Kaiser {\it et al.}~\cite{kaiser2023multiplex}. 
The DC method utilizes assortative mesoscale structure to inform its reconstruction, however, using the community structure obtained from the remnant networks. Thus, it struggles to produce valuable predictions when such structure is impossible to infer or noisy. The currently proposed method, however, uses the more flexible technique of graph embeddings to learn network structure; this, in combination with its regularization from the logistic regression, offers a reconstruction method that is noticeably more stable.

\subsubsection{Dissecting reconstruction models}
The proposed reconstruction method utilizes more than an embedded graph's node vectors alone, however, in making its classifications. In order to examine the contribution of the information provided by the graph embedding to the reconstruction procedure, we compare the magnitude of the coefficients' values learned by the logistic regression models. Specifically, we measure the magnitude of the coefficients $b$ and $c$ respectively associated to the features $\kappa$ and $\delta$ in the logistic regression model, see 
\Cref{eq:degree,eq:emb,eq:logistic},
and estimate the relative contribution of the embedding feature $\delta$ to the reconstruction of the multiplex, i.e., $|b|/(|b|+|c|)$.
Please note that the coefficient $a \simeq 0$, as these networks have approximately identical number of edges in layers $\alpha$ and $\beta$.
With these measurements, in the same controlled settings and panel ordering as \Cref{fig:perfs-LFR}, we see some immediate results. 

\begin{figure}[htb]
    \centering
    \includegraphics[width=0.47\textwidth]{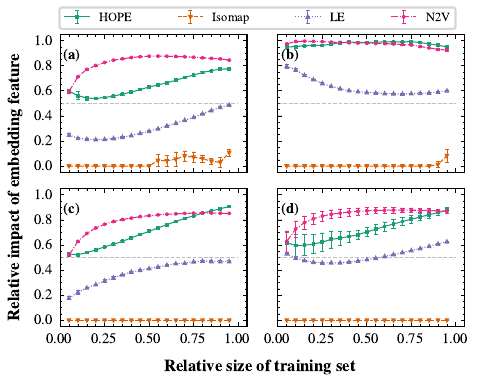}
    \caption{
        {\bf Relative impact of the embedding feature in the reconstruction of synthetic multiplex networks.} Each panel depicts the relative impact of the embedding feature in the reconstruction model, calculated as 
        $|b|/(|b|+|c|)$ with $b$ and $c$ coefficients of the logistic regression model of 
        \Cref{eq:logistic},
        as a function of training information available.
        Results are obtained on the same set of experiments as in Figure~\ref{fig:perfs-LFR}.
        All points depict the mean of ten realizations with error bars displaying the standard error. 
    }
    \label{fig:coefs-LFR}
\end{figure}

We have placed a guideline indicating an even contribution of embedding features and degree features (dashed gray horizontal line) in every panel of \Cref{fig:coefs-LFR}. From here, it takes only a moment to see that N2V and HOPE always rely on the embedding feature for the majority of the reconstruction calculations. These methods are also those which perform best according to \Cref{fig:perfs-LFR}. This is even the case in panels (c) and (d) of \Cref{fig:coefs-LFR} where community mixing is $\mu=0.5$ and $\mu=0.3$, respectively. Even in the presence of what is generally considered to be weak community structure, then, N2V and HOPE are shown to be accurately reconstructing synthetic multiplex networks while gaining several insights from these embeddings as a feature in said reconstruction.

Interestingly, the spectral embedding LE appears to approach an approximately equal contribution of embedding and degree features as the training set grows in relative size. This behavior is present in all panels; however, recall we are embedding the remnants of the multiplex. As the training set grows, the graphs being embedded shrink in total size, hence spectrum information and degree sequence information must necessarily converge, as it is known that the extremities of a graph's spectra are highly correlated with connectedness and degree information of a graph~\cite{sarkarSpectralPropertiesComplex2018}. Note that in panels (a) and (c), LE's embedding feature has a minority share in the classification model's coefficients, suggesting the degree heterogeneity has a stronger effect on the usefulness of LE embeddings than community strength. 

Isomap presents a special case as foreshadowed by its behavior in \Cref{fig:perfs-LFR}. Indeed, \Cref{fig:coefs-LFR} clearly shows a non-existent or barely noticeable contribution of the embedding feature from Isomap in reconstruction - effectively, utilizing Isomap in these scenarios was an inefficient use of only the degree feature shared by all reconstruction models. This is elaborated on further in \Cref{sec:appendix:isomap}.
Furthermore, Isomap is, in comparison, an expensive embedding method as described in \Cref{sec:appendix:time}.
Consequently, we would caution the interested researcher against using Isomap to reconstruct multiplexes with strong suspected community structure. Isomap is not without its merits, however; in real systems with more nuanced mesoscale structure, it is able to provide valuable insights, as discussed further in \Cref{sec:results:real}.

\subsubsection{Reconstruction of synthetic networks with variable community strength and degree heterogeneity}
There may be an elbow in the performance measurements of our reconstruction method, regardless of embedding method, at a relative training set size of approximately 0.3 as suggested in \Cref{fig:perfs-LFR}. While this elbow is weak in some cases, it nonetheless sets itself as a potential threshold for which we can expect diminishing performance returns beyond it with respect to returns made before this pseudo-elbow. We leverage this observation to conduct additional reconstruction experiments on benchmarks with a higher resolution than given in \Cref{fig:perfs-LFR,fig:coefs-LFR}. Specifically, we explore samples of our synthetic duplex space over $\mu \in [0.1, 0.5], \gamma \in [2.1, 3.9]$ with the same fixed size and 
degree distribution constraints as above. 

\begin{figure}[hbt]
    \centering
    \includegraphics[width=0.47\textwidth]{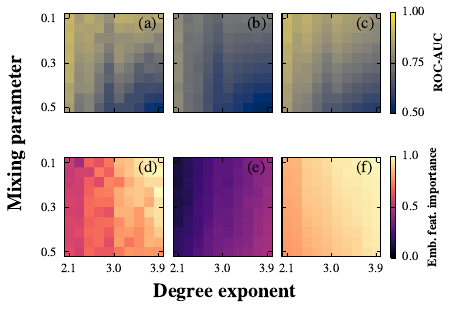}
    \caption{
        {\bf Reconstruction of synthetic multiplex networks.} In all panels, we display a heatmap of $10 \times 10$ cells as $\mu$ varies in equal steps between 0.1 and 0.5 (inclusive) along the vertical axis and $\gamma$ varies in equal steps between 2.1 and 3.9 (inclusive) along the horizontal axis. Rows separate measurements of reconstruction performance via ROC-AUC (top) and relative magnitude of the embedding vector distance feature (bottom). Columns separate embedding methods; from left to right, we report results for HOPE, LE, and N2V. That is, the panels are as follows: (a) performance with HOPE, (b) performance with LE, (c) performance with N2V, (d) embedding feature importance with HOPE, (e) embedding feature importance with LE, (f) embedding feature importance with N2V. Throughout all panels the training set is 30\% of its maximum possible size. Each cell is furthermore reporting the mean value of five repetitions.
    }
    \label{fig:heatmap-LFR}
\end{figure}

We display the average reconstruction performance and relative feature importance in \Cref{fig:heatmap-LFR}. Given the insights on reconstruction with Isomap made available in \Cref{fig:perfs-LFR,fig:coefs-LFR}, we have elected not to report it in this figure; the reader may find a similar visualization for Isomap in \Cref{sec:appendix:isomap}. 

As one may intuitively expect, reconstruction is generally easier in the presence of structural heterogeneity - that is, with strong community structure and heterogeneous degrees. With strong community structure but weakening degree heterogeneity, HOPE and N2V are generally able to yield mostly correct reconstructions; LE, however, appears to suffer from a worsened performance gradient with respect to $\gamma$. From the top row of \Cref{fig:heatmap-LFR}, one can see our prior interpretation of N2V as the most stable of the tested embeddings more clearly - namely, the average gradient throughout $\mu-\gamma$ space of ROC-AUC is smallest with N2V. 

Furthermore, from the perspective of relative feature contributions to the classification model, N2V remains the most stable method, as seen by the bottom row of \Cref{fig:heatmap-LFR} for similar reasons as above. LE utilizes its embedding feature for classifications strikingly less intensely than either HOPE or N2V, regardless of community strength or degree heterogeneity. All methods share an increasing contribution of embedding feature contributions as degree heterogeneity weakens, however.

\subsection{Real networks}\label{sec:results:real}
\subsubsection{Performance on real networks}
\begin{figure}[htb]
    \centering
    \includegraphics[width=0.47\textwidth]{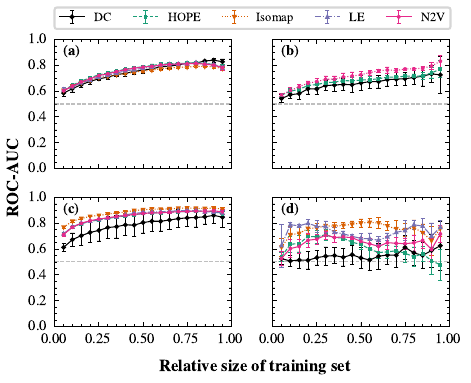}
    \caption{
        {\bf Reconstruction of real multiplex networks.} Panel (a) depicts the reconstruction of ``cond-mat.dis-nn" and ``cond-mat.stat-mech" of the arXiv collaboration network; (b) concerns electric and chemical monadic synaptic junctions in the \textit{C. Elegans} connectome; (c) concerns direct and suppressive genetic interactions in the \textit{Drosophila} genome; and (d) concerns overground trolley and Docks Light Railway transport lines in London. 
        In all panels, we plot the ROC-AUC as a function of the relative size of the training set.
        All points depict the mean of ten realizations with error bars displaying the standard error. 
    }
    \label{fig:perfs-real}
\end{figure}

While our results so far illuminate reconstruction potential in ideal scenarios, real-world systems are of course more nuanced. We analyze a collection of real-world multiplexes previously considered in related works; these are described further in \Cref{sec:data:real}. These networks host a variety of interesting structures and are drawn from several domains. As with the synthetic multiplexes, we train our reconstruction algorithm on random training sets of increasing size and report the performance of the algorithm in \Cref{fig:perfs-real}. 

We generally observe good reconstruction performance in these real-world multiplexes. In all cases, it is easily seen in \Cref{fig:perfs-real} that graph embeddings in multiplex reconstruction outperform the DC method.
Within the estimated error (standard error of the mean), N2V in particular is consistently performing better than the DC method in all cases here, except with exceptionally large training sets on the arXiv collaboration multiplex as shown in \Cref{fig:perfs-real} panel (a). The absolute error of the reconstruction performances is also significantly smaller than for the DC algorithm, suggesting that the proposed framework is able to produce more self-consistent reconstructions in a variety of systems. 

In addition to the improved stability of the proposed reconstruction framework, there is a steady improvement in performance as well. While marginal in some systems, the improvement is undeniably more significant in others. The rail transportation multiplex in London is easier to reconstruction with the proposed method than with the DC method,
as seen in \Cref{fig:perfs-real} panel (d). This multiplex has a nontrivial latent spatial embedding and very path-like topology in each layer. This topology confounds modularity maximization techniques, however, random-walk and geodesic-based embedding methods, such as N2V and Isomap, are able to uncover this topology and leverage it effectively to approximate the true multiplex structure.

\subsubsection{Dissecting reconstruction models}
\begin{figure}[htb]
    \centering
    \includegraphics[width=0.47\textwidth]{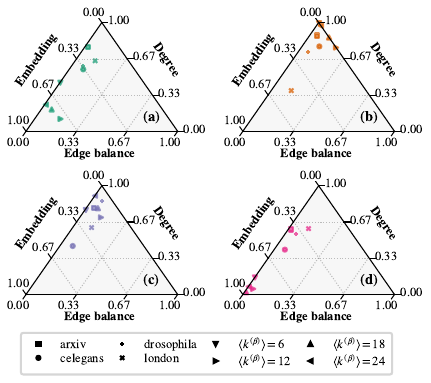}
    \caption{
        {\bf Relative impact of edge-balance, embedding, and degree features in the reconstruction of multiplex networks.} We show the model coefficients for reconstructing real systems at a fixed relative training set size of 30\%, as well as four synthetic duplexes of varying imbalances (triangles). The coefficients are displayed in a two-dimensional simplex, hence, the coefficients must add to 1 along the three dimensions.
        Values can be read by taking the line parallel to the guidelines to a given point - the intersection with the axes gives the coordinates. The intersection of guidelines in the center of the figure represents a model with equal contribution of embedding, degree, and edge-balance features.  Panel (a) reports coefficients for the HOPE embedding, (b) for Isomap, (c) for LE, and (d) for N2V. Different symbols correspond to different real systems (non-triangles) or synthetic multiplex networks (triangles) as given by the legend. For the synthetic multiplex networks, we replicate parameters as in \Cref{fig:perfs-LFR} panel (a) with the exception of the average degree of layer $\beta$, noted in the legend. Each point is the average of five repetitions.  
    }
    \label{fig:coefs-real}
\end{figure}

In a similar normalization as presented in \Cref{fig:coefs-LFR}, we display the coefficients of models applied to real systems in \Cref{fig:coefs-real}.
Noticeably, since these systems are not necessarily balanced - that is, $|\mathcal{E}^{(\alpha)}| \neq |\mathcal{E}^{(\beta)}|$ - we no longer have null intercepts. 
We report the normalized model coefficients in \Cref{fig:coefs-real}, with panel structure now delineating graph embedding and symbols showing different real-world multiplexes, as well as some synthetic multiplex networks with varying edge balance between the layers. 
Both the real-world systems and the synthetic multiplex networks display, despite varying degrees of imbalance, an edge-balance coefficient smaller in magnitude than either the embedding feature or degree feature coefficients.
This suggests that the use of these features is truly informative for multiplex reconstruction in these settings.

Consider the real-world systems (non-triangular symbols).
There are some clear differences from the synthetic cases seen in \Cref{fig:coefs-LFR}.
Perhaps most noticeable is the non-trivial contribution of Isomap in all cases as seen in \Cref{fig:coefs-real} panel (b).
Re-examining \Cref{fig:perfs-real} panel (d), we were able to leverage Isomap to reconstruct a rail transportation multiplex more faithfully than with other embeddings.
This speaks to one of the strengths of incorporating graph embeddings into multiplex reconstruction, namely, that we are able to inherit advantages of the graph embeddings themselves.
Isomap acts on shortest-path distances of the graph, extracting similarities on the basis of geodesic distance.
Since, in a transportation system such as London's rail transportation multiplex, these geodesic distances are correlated with spatial distances of the corresponding locations, Isomap is able to grasp additional information than the other methods.
This is especially clear in \Cref{fig:coefs-real} panel (b), where we see the embedding feature coefficient for London (orange x) constitutes the majority of the feature contributions.

Another stark difference from the synthetic case of \Cref{fig:coefs-LFR} can be seen with HOPE and N2V's embedding feature contributions, \Cref{fig:coefs-real} panels (a) and (d). In the synthetic case, these enjoyed a significant majority of reconstruction model contributions over degree, as seen in all panels of \Cref{fig:coefs-LFR}. However, in these real systems the contribution of these embedding features, as in all systems and embeddings, is a noticeably smaller portion. This is perhaps not entirely surprising, as real-world systems have a mesoscale structure than is less neat than that of synthetic models. This does not seem to deter from N2V and HOPE's generally good reconstruction performance, however, as can be seen from their relative closeness to the other methods' performance in \Cref{fig:perfs-real}, with the exception of London's rail transportation as discussed above. 

Regardless of embedding method, the real systems yield model coefficients with non-negligible contributions from the embedding feature. 

Now consider the imbalanced synthetic multiplex networks.
As in \Cref{fig:coefs-LFR}, we see that HOPE and N2V make extensive use of the embeddings to do the reconstruction, as can be seen in \Cref{fig:coefs-real} panels (a) and (d) (triangular symbols).
Interestingly, there does not seem to be a strong effect from the edge imbalance of the two layers, evident from the small value of the edge balance coefficient across all embedding methods.
We suspect this is due to the strong mesoscale structure and degree heterogeneity yielding informative features, lowering the utility of edge balance for reconstruction purposes.
Also similar to \Cref{fig:coefs-LFR}, panel (b) displays a generally uninformative embedding feature when embedding with Isomap. Likewise, panel (c) displays a weaker contribution of the LE-based embedding feature than HOPE or N2V for reconstructing strongly structured synthetic multiplex networks.

\section{Discussion}
We have presented a technique to reconstruct multiplex networks from their single-layer aggregates and demonstrated several notable improvements over the current state of the art. By relying on graph embeddings, we have been able to reconstruct multiplexes in a similar setting as in Refs.~\cite{kaiser2023multiplex,wu2022discrimination} but more accurately and with greater self-consistency. Furthermore, by extending the classification model into a regression framework, we have demonstrated that the proposed framework is also able to account for a variety of different mesoscale structures useful to infer the hidden multiplex structure of a network.
Given that it has been brought to light that there are dangers to modeling real-world systems as single-layer networks when they are, in reality, multiplex networks~\cite{boccalettiStructureDynamicsMultilayer2014b,bianconi2018multilayer,zaninCanWeNeglect2015},
we believe our framework could represent a valuable tool for many researchers.

Despite our presented method makes several assumptions that are not representative of all real-world systems, our results are nonetheless promising enhancements of multiplex network modeling. We are able to faithfully reconstruct several multiplexes fitting the problem statement. We make no pretense on the ubiquity of our modeling assumptions, however, which should be considered carefully by the interested practitioner. The likelihood model chosen for individual edge placements is a noticeable example of this. However, much like recent calls to action in network science ~\cite{peelStatisticalInferenceLinks2022}, this apparent short-coming is, in truth, capable of being a great strength. The researcher in need of multiplex reconstruction must be intentional in their hypotheses of the prior of multiplex structure before our reconstruction method can be adequately utilized. This places more work on the interested researcher, but not for nothing; clearer, more specific hypotheses are the basis of rigorous, reproducible science and should be embraced, not viewed as a barrier to progress. We have provided ample evidence that the baseline generative models we assume in this investigation are still informative in a variety of systems and hence that the reconstruction approach described has potential in further applications and this, in itself, is an interesting observation. 

The investigation presented here should not be considered a complete solution to the multiplex reconstruction problem. There are avenues of future research readily available to extend the developed framework for both pragmatic use in data collection and processing, and also in foundational understanding in the limitations of multidimensional network compression. Some examples of the former are generalizations of the method to directed multiplexes and treatment of multiplexes with overlapping layers. We suspect directedness to provide mathematically plain but useful insights; however, addressing overlapping layers may increase the dimensionality of the underlying classification problem in unpredictably complicated ways. 
\change{Also, considering the possibility that the training set is biased, meaning that the observed edges are not uniformly selected from the layers rather preferentially picked from one layer instead of the others, is a realistic and interesting avenue for future research.}
Lastly, further investigations on the fundamental reconstructability of multiplexes give insight into the reverse problem of multiplex aggregation as well. There are many well-done studies on compressing the layer dimension of multiplex systems and when this compression irretrievably loses information~\cite{de2015structural,ghavasiehEnhancingTransportProperties2020,ghavasiehDismantlingInformationFlow2023,zaninCanWeNeglect2015}; further work in the multiplex reconstruction problem may illuminate attempted error-correction on disaggregating compressed multiplex data.

\section{Conclusion}
We have described and systematically investigated a method of reconstructing multiplex networks from single-layer aggregate data and partial multidimensional observations.
This so-called ``multiplex reconstruction problem," while not discussed for the first time in this article, is nonetheless in its infancy and promises a rich source of inquiry in contemporary network science.
Our current work has suggested an inference framework for solving the multiplex reconstruction problem under certain conditions.
In favor of rigorous future science, the proposed method also makes improvements over the current literature in hypothesis testing and generative modeling.
As scientists continue recent discoveries in the hidden or confounded multiplexity of real-world systems, we hope our proposed methodology can serve as a valuable tool in ensuring network datasets are faithful to the systems they describe.  

\section*{Acknowledgment}
This project was partially supported by the Army Research Office under contract number W911NF-21-1-0194 and by the Air Force Office of Scientific Research under award number FA9550-21-1-0446. The funders had no role in study design, data collection and analysis, the decision to publish, or any opinions, findings, and conclusions or recommendations expressed in the manuscript.

\appendix
\section{Imbalanced Layers}\label{sec:appendix:imbalanced}

\begin{figure}[hb]
    \centering
    \includegraphics[width=0.47\textwidth]{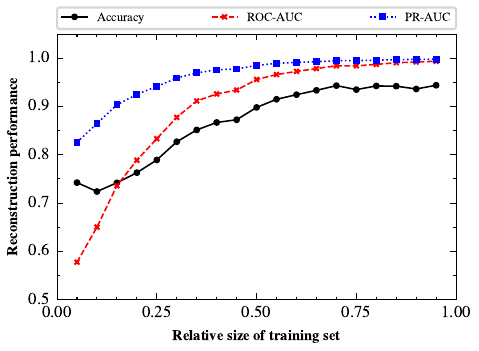}
    \caption{
        {\bf Reconstruction performance on an imbalanced synthetic duplex.} We reconstruct a synthetic duplex with layers formed as described in \Cref{sec:data:synth}, albeit where one layer has average degree 6 and the other has average degree 20. The test edges are approximately twice as likely to have originated from the denser layer than the sparser $\Big(\frac{|\mathcal{E}^{(\alpha)}|}{|\mathcal{E}^{(\beta)}|} \approx 0.3229 \Rightarrow \zeta \approx -0.5118 \Big)$. We display the accuracy, area under the receiver operating characteristic, and area under the precision-recall curve as different perspectives on the reconstruction's performance. 
    }
    \label{fig:imbalanced}
\end{figure}

Many real-world multiplexes have layers with widely varying sizes, as measured by the number of their edges. It is well known that classification models can be mislead by imbalanced classes; to ensure we have not been mislead by class imbalance in our own classification setting, we explored reconstruction of duplex models with highly imbalanced layers under common metrics of a binary classifier's performance. Precision-recall is known to be preferred to ROC-AUC in imbalanced classification tasks~\cite{saitoPrecisionRecallPlotMore2015} and indeed we confirm this in the reconstruction setting of \Cref{fig:imbalanced}. We also observe, however, that ROC-AUC, while strongly affected by imbalance, is nonetheless monotonic as a function of the size of the training set. 

\section{Workflow time complexity}\label{sec:appendix:time}

\begin{figure}[htbp]
    \centering
    \includegraphics[width=0.47\textwidth]{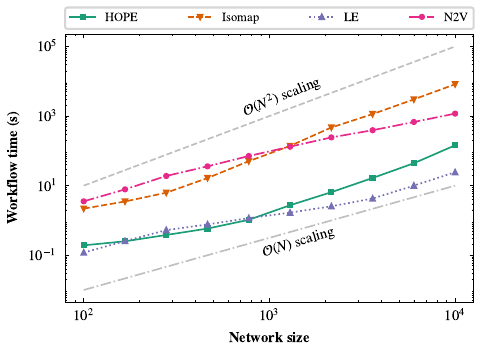}
    \caption{
        {\bf Algorithmic complexity of reconstruction workflow.} We generate synthetic networks as in \Cref{sec:data:synth} with varying sizes but fixed $\gamma=2.1, \tau=1.0, \langle k^{(\alpha)} \rangle = \langle k^{(\beta)} \rangle = 6, \text{ and } k_{\text{max}} = 100$. Each point is the average of 10 repetitions. Error
        bars of the standard error are obscured by markers at this scale and removed from the figure. Workflow complexity is dominated by the embedding method ranging between linear and quadratic scaling for these methods. Simulations were run on an Intel(R) Xeon(R) CPU E5-2690 v4 @ 2.60 GHz. All simulations were restricted to a single core.
    }
    \label{fig:time-complexity}
\end{figure}

The reconstruction workflow we have presented here consists of several independent calculations on the layers of the induced remnant multiplex. One may rightly be concerned with the computational complexity of multiplex reconstruction in this framework. The dominating complexity of the workflow, however, is that of the selected graph embedding method. The reconstruction workflow is no more expensive, asymptotically, than the selected graph embedding method - of which, considerable effort has been placed into quantifying and optimizing. According to a recent review ~\cite{zhang2021systematic}, graph embedding complexities for these methods can be summarized as in \Cref{tab:complexities}. We show the total reconstruction workflow time on synthetic duplexes of varying sizes in \Cref{fig:time-complexity}.

\begin{table}[b]
    \centering
    \begin{tabular}{|l|r|}
        \hline
        Embedding & Time complexity \\
        \hline
        \hline
        HOPE & $\mathcal{O}(d^2E)$\\
        Isomap & $\mathcal{O}(CN^2 + dN^2)$\\
        LE & $\mathcal{O}(d^2E)$\\
        N2V & $\mathcal{O}(dN)$\\
        \hline
    \end{tabular}
    \caption{
        {\bf Asymptotic time complexity of graph embeddings.} $E$ represents the total number of edges in the graph, $d$ represents the embedding dimension, and $C$ represents the cost to compute the shortest paths between any pair of nodes.
    }
    \label{tab:complexities}
\end{table}

\section{Size effects}\label{sec:appendix:size}

\begin{figure}[ht]
    \centering
    \includegraphics[width=0.47\textwidth]{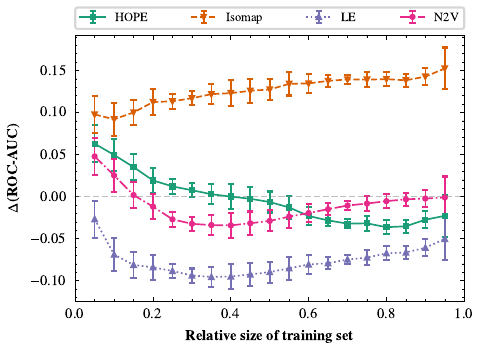}
    \caption{
        {\bf Reconstruction of synthetic multiplex networks of different sizes.} Using the same synthetic model as throughout the rest of this document, albeit with $N \in \{1000, 10000 \}$, we show the difference of the average ROC-AUCs as the training set grows. The dashed gray line represents no performance difference in average performances. We see the absolute performance of N2V and HOPE's performance is relatively small, potentially negligible. LE and Isomap, however, display robust, noticeably worse/better performance with $N=10000$ than $N=1000$, respectively. Each point depicts the difference of ten repetitions in each network size, with standard error given by the error bars.
    }
    \label{fig:size-effect}
\end{figure}

Synthetic multiplex networks of a fixed size are investigated in \Cref{fig:perfs-LFR,fig:coefs-LFR,fig:heatmap-LFR}. We fixed the size in these analysis so as to better facilitate comparison in other variables of interest. Now, however, we turn to examine the effect on reconstruction performance the size of the multiplex may have as illustrated in \Cref{fig:size-effect}. 

In particular, we show the difference of average ROC-AUCs for synthetic models with 10,000 and 1000 nodes. From the scale of the y-axis, it is clear that N2V and HOPE have differences as network size increases at these scales, but Isomap and LE display yet more noticeable differences. This may not strike the reader as unexpected; Isomap in particular is an embedding of geodesic distances and even under normalization we would expect increasing network size to non-trivially affect the reconstruction. LE's reconstructions get less performant as network size increases.

\section{Isomap on LFRs}\label{sec:appendix:isomap}

We conducted a sweep over $\mu-\gamma$ space using Isomap as in \Cref{fig:heatmap-LFR}. Isomap, however, does 
a
poor job in reconstructing multiplexes as tested here. 

\begin{figure}[!hb]
    \centering
    \includegraphics[width=0.47\textwidth]{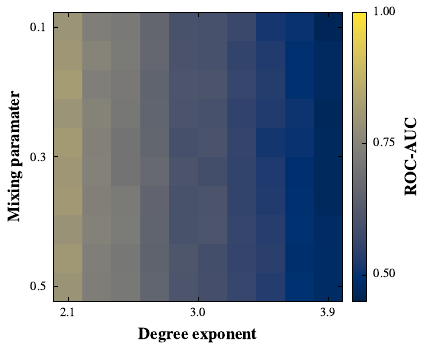}
    \caption{
        {\bf Reconstruction performance on synthetic multiplex networks with Isomap.} We display the equivalent of the ROC-AUC heatmaps of \Cref{fig:heatmap-LFR} albeit with Isomap as the graph embedding. The embedding feature importances for Isomap here are all near zero and indistinguishable with most colormaps, hence that corresponding heatmap for Isomap is not shown.
        }
    \label{fig:heatmap-LFR-Isomap}
\end{figure}

This is made clearer by observation that the color gradient in \Cref{fig:heatmap-LFR-Isomap} differentiated primarily, up to random noise, horizontally. The average relative importance of the Isomap embedding feature in all cases was near 0, hence, we have not included its corresponding heatmap.

\nocite{*}

%

\end{document}